\documentclass[12pt]{article}

\usepackage{graphicx}
\usepackage{amsmath,amssymb}
\usepackage{bm}
\usepackage{graphicx, color}
\begin{document}
\title{
\begin{flushright}
\ \\*[-80pt] 
\begin{minipage}{0.2\linewidth}
\normalsize
\end{minipage}
\end{flushright}
{\Large \bf 
Deviation from  tri-bimaximal mixing  and  flavor  symmetry breaking 
 in a seesaw type  $A_4$  model 
\\*[20pt]}}

\author{
\centerline{
Atsushi~Hayakawa$^{1,}$\footnote{E-mail address: hayakawa@muse.sc.niigata-u.ac.jp},   
Hajime~Ishimori$^{1,}$\footnote{E-mail address: ishimori@muse.sc.niigata-u.ac.jp}, } \\ 
\centerline{
~Yusuke~Shimizu$^{1,}$\footnote{E-mail address: shimizu@muse.sc.niigata-u.ac.jp},
Morimitsu~Tanimoto$^{2,}$\footnote{E-mail address: tanimoto@muse.sc.niigata-u.ac.jp} }
\\*[20pt]
\centerline{
\begin{minipage}{\linewidth}
\begin{center}
$^1${\it \normalsize
Graduate~School~of~Science~and~Technology,~Niigata~University, \\ 
Niigata~950-2181,~Japan } \\
$^2${\it \normalsize
Department of Physics, Niigata University,~Niigata 950-2181, Japan } 
\end{center}
\end{minipage}}
\\*[50pt]}

\date{
\centerline{\small \bf Abstract}
\begin{minipage}{0.9\linewidth}
\medskip 
\medskip 
\small
We have studied  the contribution of 
 higher order corrections of the flavor symmetry breaking
in the $A_4$ seesaw model with the supersymmetry.
 Taking account of  possible  higher dimensional mass operators, 
we  predict  the deviation from the tri-bimaximal lepton mixing  for both
normal hierarchy and inverted hierarchy of  neutrino masses.
We have found that the value of $\sin^2 2\theta_{23}$ is larger than $0.96$
and the upper bound of $\sin^2\theta_{13}$ is $0.01$.
  We have also examined the flavor changing neutral current of leptons
 from the soft SUSY breaking  in 
slepton masses and A-terms within 
the framework of supergravity theory. Those magnitudes
are  enough suppressed to be consistent with experimental constraints.
\end{minipage}
}

\begin{titlepage}
\maketitle
\thispagestyle{empty}
\end{titlepage}

\section{Introduction}
Lepton flavor mixing provides us an important clues to understand 
 the origin of the generation. 
Recent neutrino oscillation experimental data \cite{Threeflavors,fogli}
 indicate the  tri-bimaximal mixing  for three lepton flavors \cite{HPS}. 
Indeed, various types of models leading to the tri-bimaximal mixing 
have been proposed, e.g. by assuming 
several types of non-Abelian flavor symmetries.
In particular, natural models realizing the tri-bimaximal mixing  
have  been proposed  based on the non-Abelian finite group $A_4$
\cite{A4}-\cite{A4-Ghosal}.
Since neutrino experiments  go into the new  phase  
of precise  determination of mixing angles and mass squared  differences, 
 it is important to study   the $A_4$ flavor model in detail.

The $A_4$ flavor model  considered by Alterelli et al \cite{Alta1,Alta2},
which  realizes  the tri-bimaximal flavor mixing, can  predict 
the deviation from the tri-bimaximal mixing.
  Actually, one of authors has  investigated
the deviation from the tri-bimaximal mixing including
 higher dimensional operators  in the effective model 
without right-handed Majorana neutrinos \cite{A4-Honda}.
In that paper, the effect on  the alignment of vacuum
 from   higher dimensional operators  was taken  account numerically.

In present paper, we discuss the $A_4$ flavor model with the supersymmetry
 including the right-handed neutrinos.
We take into account higher dimensional  operators of  neutrino masses
in  the seesaw model, and then  
  predict the deviation from the tri-bimaximal mixing.
It is found that this deviation is dominated by the vacuum expectation 
value of $\phi_{T_1}$, which is the first component of an $A_4$ triplet scalar. 
Since  the  vacuum alignment is an important ingredient to reproduce
 the tri-bimaximal mixing of neutrinos,
the effect of the shift  of  the  vacuum alignment
due to  higher dimensional operators is also discussed.
This effect is found to be negligibly small.

On the other hand, although squarks and sleptons have not been detected yet, 
their mass matrices are strongly constrained by 
experiments of flavor changing neutral current (FCNC) processes.
Non-Abelian flavor symmetries and certain types of 
their breaking patterns are useful to suppress FCNCs.
(See e.g. \cite{Kobayashi:2003fh,Ko:2007dz,Ishimori:2008ns,A4soft}.) 
In addition to flavor symmetries, their breaking 
patterns are important to derive lepton mass matrices 
and to predict  slepton mass matrices.
Therefore, we study which pattern of slepton mass 
matrices is predicted from the seesaw type  $A_4$ flavor model 
including  higher dimensional  operators and 
to examine whether the predicted pattern of slepton mass 
matrices is consistent with the current FCNC experimental bounds
\footnote{We have studied  soft 
supersymmetry (SUSY) breaking terms of sleptons  in the $A_4$ flavor model 
without right-handed Majorana neutrinos \cite{A4soft}.
}.

In Section 2, we present the lepton superpotential including
  higher dimensional operators in the $A_4$ model~\cite{Alta2}.
 We discuss the  charged lepton mass matrix and the neutrino
mass matrix in section 3.
In section 4, the lepton mixing matrix is studied to find the deviation
from the tri-bimaximal mixing matrix numerically.
In Section 5, we discuss  soft supersymmetry (SUSY) breaking terms of 
sleptons, i.e. soft scalar mass matrices and A-terms.
Section 6 is devoted to the summary.

\section{Lepton superpotential}

We begin by discussing the supersymmetric seesaw type $A_4$ flavor model  
proposed by Alterelli et al \cite{Alta1,Alta2}.
 In the non-Abelian finite group $A_4$, there are twelve group elements 
and four irreducible representations: $1$, $1'$, $1''$ and $3$. 
The $A_4$ and $Z_3$ charge assignments of leptons and scalars are
listed in Table 1. Under the 
$A_4$ symmetry, the chiral superfields for three families of 
the left-handed lepton doublet 
$l=(l_e,l_\mu,l_\tau)$ and right handed neutrino $\nu ^c=(\nu_e^c,\nu_\mu^c,\nu_\tau^c)$ are assumed to transform as 3, 
while the right-handed ones of the charged 
lepton singlets $e^c$, $\mu^c$ and $\tau^c$ are assigned with $1$, $1''$, $1'$, respectively. 
The third row of Table 1 shows how each chiral multiplet transforms 
under $Z_3$, where $\omega= e^{2\pi i/3}$. 
The flavor symmetry is spontaneously broken by vacuum expectation values (VEV) of two $3$'s, $\phi_T$, $\phi_S$, and by one singlet, $\xi$, 
which are $SU(2)_L\times U(1)_Y$ singlets. 
Their $Z_3$ charges are also shown in Table 1. 
Hereafter, we follow 
the convention that the chiral 
superfield and its lowest component are denoted by the same letter. 

\begin{table}[tbh]
\begin{footnotesize}
\begin{tabular}{|c|ccccc||cc||ccccc|}
\hline
              &$(l_e,l_\mu,l_\tau)$ & $(\nu_e^c,\nu_\mu^c,\nu_\tau^c)$ & $e^c$ & $\mu^c$ & $\tau^c$ & $h_u$ & $h_d$ &$ \xi $ &$ \tilde \xi $&  $(\phi_{T_1},\phi_{T_2},\phi_{T_3})$  &  $(\phi_{S_1},\phi_{S_2},\phi_{S_3})$ &$\Phi$ \\ \hline
$A_4$      &$3$  &$3$    &$1$ &$1''$&$1'$  &$1$ &$1$&$1$&$1$& $3$ & $3$  &$1$ \\
$Z_3$      &$\omega$  &$\omega^2$    &$\omega^2$ &$\omega^2$&$\omega^2$  &$1$ &$1$&$\omega^2$&$\omega^2$& $1$ & $\omega^2$  &$1$ \\
$U(1)_{FN}$      &0  &0    &$2q$ &$q$&0  &0 &0&0&0& 0 & 0  &$-1$ \\
\hline
\end{tabular}
\end{footnotesize}
\caption{ $A_4$, $Z_3$ and $U(1)_{FN}$ charges}
\end{table}

Allowed terms in the superpotential including charged leptons are 
written by
\begin{eqnarray}
w_l
&=&
y_0^e 
  e^c  l\phi_Th_d\frac{\Phi^{2q}}{\Lambda'^{2q}}\frac{1}{\Lambda}
+y_0^\mu
  \mu^c  l\phi_Th_d\frac{\Phi^{q}}{\Lambda'^{q}}\frac{1}{\Lambda}
+y_0^\tau 
  \tau^c  l\phi_Th_d\frac{1}{\Lambda}
\nonumber\\&&
+y_1^e
 e^c l\phi_T\phi_Th_d\frac{\Phi^{2q}}{\Lambda'^{2q}}\frac{1}{\Lambda}
+y_1^\mu
 \mu^c l\phi_T\phi_Th_d\frac{\Phi^{q}}{\Lambda'^{q}}\frac{1}{\Lambda}
 \nonumber\\&&
+y_1^\tau
 \tau^c l\phi_T\phi_Th_d\frac{1}{\Lambda}. 
\label{charged}
\end{eqnarray}
In our notation, all $y$ with some subscript denote Yukawa couplings 
of order $1$ and $\Lambda$ denotes cut off scale of the $A_4$ symmetry. 
In order to obtain the natural hierarchy among lepton masses $m_e$, $m_\mu$ 
and $m_\tau$, the Froggatt-Nielsen mechanism \cite{FN} is introduced 
as an additional  $U(1)_{FN}$ flavor symmetry 
under which only the right-handed lepton sector is charged. 
$\Lambda'$ is  a cut off scale of the $U(1)_{FN}$ symmetry and
$\Phi$ denotes the Froggatt-Nielsen flavon in Table 1.
The $U(1)_{FN}$ charge values are 
taken as $2q$, $q$ and 0 for $e^c$, $\mu^c$ and $\tau^c$, respectively. 
By assuming that a flavon, 
carrying a negative unit charge of $U(1)_{FN}$, acquires a VEV 
$\left<\Phi\right>/\Lambda'\equiv\lambda \ll 1$, the following 
mass ratio is  realized through the Froggatt-Nielsen charges, 
\begin{eqnarray}
 m_e :  m_\mu:  m_\tau = \lambda^{2q} : \lambda^q :1.
\end{eqnarray}
If we take  $q = 2$,  $\lambda\sim 0.2$ is required to be consistent with
the observed charged lepton  mass hierarchy. 
The $U(1)_{FN}$ charges are listed  in the fourth row of Table 1.

The superpotential associated with 
 the Dirac neutrino mass is given as
\begin{eqnarray}
w_D
&=&
y_0^D \nu^c lh_u 
+y_1^D \nu^c lh_u\phi_T\frac{1}{\Lambda},
\label{dirac}
\end{eqnarray}
and for the right-handed Majorana sector, the superpotential  is given as
\begin{eqnarray}
w_N
&=&
y_0^N \nu^c \nu^c \phi_S
+y_1^N \nu^c \nu^c \xi
\nonumber\\&&
+y_2^N \nu^c \nu^c \phi_T\xi\frac{1}{\Lambda}
+y_3^N \nu^c \nu^c \phi_T\phi_S\frac{1}{\Lambda},
\label{majorana}
\end{eqnarray}
where there appear $3\times3\times3$ and $3\times3\times3\times3$ 
products of $A_4$ triplets.

Vacuum alignments of $A_4$ triplet $\phi_T$ and $\phi_S$
are   required to reproduce the tri-bimaximal mixing.
These  vacuum alignments are   realized in the scalar potential 
of  the leading order \cite{Alta2}.  
 However,  higher order operators shift these  vacuum alignments,
therefore we write  vacuum expectation values (VEVs) as follows:
\begin{eqnarray}
&&\left<h_{u}\right>=v_{u},
\qquad
\left<h_{d}\right>=v_{d},
\qquad
\left<\xi\right>=u,
\nonumber\\
&&\left<(\phi_{T_1},\phi_{T_2},\phi_{T_3})\right>=v_T(1,\epsilon_1,\epsilon_2), \qquad 
\left<(\phi_{S_1},\phi_{S_2},\phi_{S_3})\right>=v_S(1,1+\delta_1,1+\delta_2),
\end{eqnarray}
where $\delta_i\ll 1$ and $\epsilon_i\ll 1$. 
The parameters $\epsilon_i$ and $\delta_i$ are  given 
in  the  model   of \cite{Alta2} as
\begin{eqnarray}
\label{epdel}
\epsilon_1=\epsilon_2 = C_0\frac{u^3}{v_T^2}\frac{1}{\Lambda }, 
\qquad
\delta_1=C_1\frac{u^3}{v_T^2}\frac{1}{\Lambda },
 \qquad
\delta_2=C_2\frac{u^3}{v_T^2}\frac{1}{\Lambda },
\end{eqnarray}
where $C_i$s are coefficients of order one.
We will estimate magnitudes of $\epsilon_i$ and $\delta_i$
in following  numerical calculations.

\section{Lepton mass matrices in  $A_4$ flavor model}
Inserting VEVs in  the superpotential of the charged lepton sector 
in Eq.(\ref{charged}), 
we obtain the charged lepton mass matrix $M_E$ as
\begin{eqnarray}
M_E
=\alpha_T v_d
\begin{pmatrix}
      y_0^e\lambda^{2q} + \frac{2}{3}y_1^e\lambda^{2q}\alpha_T &   y_0^e\lambda^{2q} \epsilon_2   &  y_0^e\lambda^{2q} \epsilon_1 \\
     y_0^\mu\lambda^q \epsilon_1 &   y_0^\mu\lambda^q +\frac{2}{3} y_1^\mu\lambda^q \alpha_T   &  y_0^\mu\lambda^q \epsilon_2 \\
    y_0^\tau \epsilon_2 &  y_0^\tau \epsilon_1 &   y_0^\tau+\frac{2}{3} y_1^\tau \alpha_T
\end{pmatrix} +\  {\cal O}(\alpha_T^2\epsilon_i v_d)\ ,
\label{chargedlepton}
\end{eqnarray}
with 
\begin{equation}
\alpha_T =\frac{v_T}{\Lambda }.
\end{equation}
In this mass matrix, the off diagonal elements
appear in order of  $\epsilon_i$.
Since we have 
\begin{eqnarray}
m_e^2&=&{y_0^e}^2\lambda^{4q} 
\alpha_T^2(1-\epsilon_1^2-\epsilon_1\epsilon_2-\epsilon_2^2)  v_d^2 ,
\nonumber\\
m_\mu^2&=&{y_0^\mu}^2\lambda^{2q} \alpha_T^2(1-2\epsilon_1\epsilon_2)  v_d^2,
\nonumber\\
m_\tau^2&=&{y_0^\tau}^2 \alpha_T^2 (1+\epsilon_1^2+\epsilon_2^2) v_d^2,
\end{eqnarray}
we can determine $\alpha_T$ from the tau lepton mass by fixing $y_0^\tau$: 
\begin{eqnarray}
\alpha_T
=\sqrt{\frac{m_\tau^2}{{y_0^\tau}^2v_d^2(1+\epsilon_1^2+\epsilon_2^2)}}.
\end{eqnarray}

Since off diagonal elements of the charged lepton mass matrix are of order $\epsilon_i$, 
the  mixing is  expected to be small.  The mixing matrix  is given as
\begin{eqnarray}
V_E
= \begin{pmatrix}1   & \theta_{12}^e & \epsilon_2 \\ 
                   -\theta_{12}^e    & 1   &\epsilon_1    \\
                   -\epsilon_2  & -\epsilon_1 & 1    \\
 \end{pmatrix},
\end{eqnarray}
the mixing angle $\theta_{12}^e$  depends on  the relative magnitude of 
$\lambda^{2q}$ and $\epsilon_i$ as 
\begin{eqnarray}
\theta_{12}^e
=\frac{{y_0^\mu}^2\lambda^{2q}+\frac13{y_0^\tau}^2\epsilon_2}
{{y_0^\mu}^2\lambda^{2q}+{y_0^\tau}^2(\epsilon_1^2-\epsilon_2^2)
-\frac49{y_1^\tau}^2\alpha_T^2} \ \epsilon_1.
\end{eqnarray}

Now, we present the Dirac neutrino mass matrix as follows:
\begin{eqnarray}
\label{dirac}
M_D
&=&v_u
\begin{pmatrix}
    y_0^D + \frac{2}{3}  y_1^D \alpha_T & 0 & 0 \\
  0 &   y_0^D -\frac{1}{3}  y_1^D \alpha_T-\frac12{y_2}^D\alpha_T & 0\\
  0& 0 &   y_0^D -\frac{1}{3}  y_1^D \alpha_T+\frac12{y_2}^D\alpha_T
\end{pmatrix},
\end{eqnarray}
where  $O(\alpha_T^2)$ terms are  neglected.
It is remarked that higher order terms come from 
$\left<\phi_{T_1}\right>$, which dominates leading terms of the
 charged lepton mass matrix in Eq.(\ref{chargedlepton}).
In the same approximation, the right-handed Majorana mass matrix is
\begin{eqnarray}
\label{majorana}
&&M_N
=2\Lambda 
\begin{pmatrix}
   \frac{2}{3}  y_0^N\alpha_S+ y_1^N\alpha_V & -\frac{1}{3}  y_0^N\alpha_S (1 + \delta_1) & -\frac{1}{3} y_0^N \alpha_S(1 + \delta_2) \\
   -\frac{1}{3}  y_0^N\alpha_S (1 + \delta_1) & \frac{2}{3} y_0^N \alpha_S(1 + \delta_2) & -\frac{1}{3} y_0^N\alpha_S+y_1^N\alpha_V  \\
   -\frac{1}{3} y_0^N \alpha_S(1 + \delta_2) & -\frac{1}{3} y_0^N \alpha_S+y_1^N\alpha_V & \frac{2}{3} y_0^N \alpha_S(1 + \delta_1)
\end{pmatrix}
\nonumber\\&&
+2\alpha_T  \Lambda 
\begin{pmatrix}
    y_{31}^N \alpha_S+ \frac{4}{9} y_{34}^N \alpha_S+\frac{2}{3}y_2^N \alpha_V   &  y_{33}^N \alpha_S+ \frac{1}{9} y_{34}^N\alpha_S -\frac{1}{6} y_{35}^N\alpha_S  & y_{32}^N \alpha_S+ \frac{1}{9} y_{34}^N\alpha_S-\frac{1}{6} y_{35}^N\alpha_S   \\
  y_{33}^N \alpha_S+ \frac{1}{9} y_{34}^N\alpha_S -\frac{1}{6} y_{35}^N\alpha_S  & y_{32}^N\alpha_S - \frac{2}{9} y_{34}^N\alpha_S +\frac{1}{3} y_{35}^N\alpha_S  &y_{31}^N\alpha_S - \frac{2}{9} y_{34}^N \alpha_S -\frac{1}{3} y_2^N \alpha_V   \\
   y_{32}^N \alpha_S+ \frac{1}{9} y_{34}^N\alpha_S -\frac{1}{6} y_{35}^N\alpha_S & y_{31}^N\alpha_S - \frac{2}{9} y_{34}^N\alpha_S-\frac{1}{3} y_2^N\alpha_V    & y_{33}^N \alpha_S- \frac{2}{9} y_{34}^N \alpha_S +
\frac{1}{3} y_{35}^N\alpha_S
\end{pmatrix},
\nonumber\\
\end{eqnarray}
where
\begin{equation}
\alpha _S=\frac{v_S}{\Lambda },\qquad \alpha _V=\frac{u}{\Lambda }.
\end{equation}
By the seesaw mechanism $M_D^TM_R^{-1}M_D$, 
we get  the neutrino mass matrix $M_\nu$, which is rather complicated.
We only display leading matrix elements 
which correspond to the neutrino mass matrix in \cite{Alta2}:

\begin{eqnarray}
\begin{split}
M_\nu
&=\frac13
 \begin{pmatrix}A+2B   & A-B & A-B \\ 
                   A-B    & A+\frac12B+\frac32 C  &A+\frac12B-\frac32C    \\
                   A-B  &a+\frac12B-\frac32C  & A+\frac12B+\frac32C  \\
 \end{pmatrix}+\cdots
\\
& = \frac{B+C}{2}\begin{pmatrix}
                                  1 & 0 & 0 \\
                                  0 & 1 & 0 \\
                                  0 & 0 & 1
                             \end{pmatrix} + \frac{A-B}{3}\begin{pmatrix}
                                                   1 & 1 & 1 \\
                                                   1 & 1 & 1 \\
                                                   1 & 1 & 1
                \end{pmatrix} + \frac{B-C}{2}\begin{pmatrix}
                                 1 & 0 & 0 \\
                                 0 & 0 & 1 \\
                                 0 & 1 & 0
                                            \end{pmatrix}+\cdots \ ,
\end{split}
\end{eqnarray}
where
\begin{eqnarray}
\begin{split}
A&=k_0({y_0^{N}}^2\alpha_S^2-{y_1^{N}}^2\alpha_V^2),
\quad
B=k_0(y_0^{N}y_1^{N}\alpha_S\alpha_V-{y_1^{N}}^2\alpha_V^2),
\quad
C=k_0(y_0^{N}y_1^{N}\alpha_S\alpha_V+{y_1^{N}}^2\alpha_V^2), \\
k_0&=\frac{{y_0^{D}}^2v_u^2}{({y_0^{N}}^2y_1^N\alpha_V\alpha_S^2-{y_1^{N}}^3\alpha_V^3)\Lambda}\ .
\end{split}
\end{eqnarray}
At the leading order, 
neutrino masses are given as $m_1=B$, $m_2=A$, and $m_3=C$. 

Our neutrino mass  matrix is no more diagonalized 
by the tri-bimaximal mixing matrix
$U_{\rm tri}$,  
\begin{eqnarray}
U_{\rm tri}
= \begin{pmatrix}2/\sqrt6   & 1/\sqrt3 & 0 \\ 
                   -1/\sqrt6   & 1/\sqrt3   &-1/\sqrt2    \\
                   -1/\sqrt6  & 1/\sqrt3  & 1/\sqrt2    \\
 \end{pmatrix}.
\end{eqnarray}
After rotating $M_\nu$ as $U_{\rm tri}^TM_\nu U_{\rm tri}$, 
diagonal components are
\begin{eqnarray}
\label{diagonal}
&&(1,1):
\frac{{y_0^D}^2v_u^2}
{2(y_0^N\alpha_S+y_1^N\alpha_V)\Lambda}
(1 + {\cal O}(\alpha_S, \alpha_V, \epsilon_i, \delta_i)) \ ,
\nonumber\\
&&(2,2):
\frac{{y_0^D}^2v_u^2}{2y_1^N\alpha_V\Lambda}
(1 + {\cal O}(\alpha_S, \alpha_V, \epsilon_i, \delta_i)) \ ,
\nonumber\\
&&(3,3):
\frac{{y_0^D}^2v_u^2}{2(y_0^N\alpha_S-y_1^N\alpha_V)\Lambda}
(1 + {\cal O}(\alpha_S, \alpha_V, \epsilon_i, \delta_i)) \ .
\label{diagonal}
\end{eqnarray}
Off diagonal elements are given as 
\begin{eqnarray}
&(1,2):&
\frac{y_0^D(2\alpha_T\alpha_V(2y_1^Dy_1^N-y_0^Dy_2^N)
+y_0^Dy_0^N\alpha_S(\delta_1+\delta_2)
+(2y_1^Dy_0^N-2y_0^Dy_{34}^N+y_0^Dy_{35}^N)\alpha_S\alpha_T)}{
6\sqrt2y_1^N\alpha_V({y_0^N}\alpha_S+{y_1^N}\alpha_V)\Lambda}v_u^2,
\nonumber\\
&(1,3):&
\frac{\sqrt3y_0^D\alpha_S\alpha_T(-2y_2^Dy_0^N-3y_0^Dy_{32}^N+3y_0^Dy_{33}^N)}
{12({y_0^N}^2\alpha_S^2-{y_1^N}^2\alpha_V^2)\Lambda}v_u^2,
\nonumber\\
&(2,3):&
\frac{y_0^Dy_0^N\alpha_S(y_0^D\delta_2-y_0^D\delta_1+y_2^D\alpha_T)}
{2\sqrt6y_1^N\alpha_V({y_0^N}\alpha_S-{y_1^N}\alpha_V)\Lambda}v_u^2,
\end{eqnarray}
which are suppressed in ${\cal O}(\alpha_T,\alpha_V,\delta_i)$ compared with  diagonal elements.
Therefore, mass eigenvalues are almost determined  by Eq.(\ref{diagonal}).
On the other hand, we can evaluate the deviation from the tri-bimaximal mixing 
from the neutrino sector:
\begin{eqnarray}
&&\theta_{12}^\nu\approx
\frac{2\alpha_T\alpha_V(2y_1^Dy_1^N-y_0^Dy_2^N)
+y_0^Dy_0^N\alpha_S(\delta_1+\delta_2)
+(2y_1^Dy_0^N-2y_0^Dy_{34}^N+y_0^Dy_{35}^N)\alpha_S\alpha_T}
{3\sqrt2y_0^Ny_0^D\alpha_S},
\nonumber\\
&&\theta_{13}^\nu\approx
-\frac{\alpha_S\alpha_T(2y_2^Dy_0^N+3y_0^Dy_{32}^N-3y_0^Dy_{33}^N)}
{4\sqrt3y_1^N{y_0^D}\alpha_V},
\nonumber\\
&&\theta_{23}^\nu\approx
\frac{y_0^N\alpha_S(y_0^D\delta_1-y_0^D\delta_2-y_2^D\alpha_T)}
{\sqrt6{y_0^D}(y_0^N\alpha_S-2y_1^N\alpha_V)}.
\end{eqnarray}

Let us estimate   magnitudes of $\alpha_S$ and $\alpha_V$.
The squared mass differences are given by using Eq.(\ref{diagonal}) as,
\begin{eqnarray}
\Delta m_{\rm atm}^2\simeq \pm\frac{(y_0^Dv_u)^4}{\Lambda^2}
\frac{y_0^N y_1^N \alpha_S\alpha_V}
{[(y_0^N\alpha_S)^2-(y_1^N\alpha_V)^2]^2}, \quad
\Delta m_{\rm sol}^2\simeq\frac{(y_0^Nv_u)^4}{4\Lambda^2}
\frac{y_0^N\alpha_S(y_0^N\alpha_S+2y_1^N\alpha_V)}
{(y_1^N\alpha_V)^2(y_0^N\alpha_S+y_1^N\alpha_V)^2},
\label{dm2}
\end{eqnarray}
where the sign $+(-)$ in $\Delta m_{\rm atm}^2$ corresponds to
 the normal (inverted)  mass hierarchy.
We can obtain  $\alpha_S$ and $\alpha_V$
from these   equations.
In the case of the normal mass hierarchy, putting 
\begin{equation}
\alpha _S = k\ \alpha _V \quad (k > 0)\ ,
\label{k1}
\end{equation}
 we have 
\begin{align}
\Delta m_\text{atm}^2 \simeq 
\frac{(y_0^Dv_u)^4}{\alpha _V^2\Lambda ^2}\frac{y_0^Ny_1^Nk}{(y_0^Nk+y_1^N)^2(y_0^Nk-y_1^N)^2}, \qquad
\Delta m_\text{sol}^2 \simeq 
\frac{(y_0^Dv_u)^4}{4\alpha _V^2\Lambda ^2}\frac{y_0^Nk(y_0^Nk+2y_1^N)}
{{y_1^N}^2(y_0^Nk+y_1^N)^2}.
\end{align}
The ratio of $\Delta m_\text{atm}^2$ and $\Delta m_\text{sol}^2$ 
is expressed in terms of $k$ and Yukawa couplings as
\begin{equation}
\frac{\Delta m_\text{atm}^2}{\Delta m_\text{sol}^2} 
\simeq \frac{4{(y_1^N)}^3}{(y_0^Nk+2y_1^N)(y_0^Nk-y_1^N)^2} \ .
\label{k}
\end{equation}
Yukawa couplings are expected to be order one since there is no symmetry
to suppress them. Then, by using Eq.(\ref{k}), we get
\begin{eqnarray}
k\simeq 1\pm \frac{2}{\sqrt{3}}
\sqrt{\frac{\Delta m_\text{sol}^2}{\Delta m_\text{atm}^2}}\simeq 1.2, \ \text{or}\ 0.8\ .
\label{k2}
\end{eqnarray}
Thus, $k$ is also expected to be order one,
that is to say, $\alpha_S\sim \alpha_V$, which indicates that 
 symmetry breaking scales of   $\xi$ and  
$\phi_{S}$ are same order  in the neutrino sector.
In the following numerical analyses, we take $k=1/3 \sim 3$.

 We also obtain a typical value:
\begin{eqnarray}
\alpha_V\sim 5.8 \times10^{-4},
\label{normalalpha}
\end{eqnarray}
 where we put   $\Lambda=2.4\times10^{18}{\rm GeV}$,
$\Delta m^2_{\rm atm}\sim 2.4\times 10^{-3}{\rm eV}^2$, 
$\Delta m^2_{\rm sol}\sim 8.0\times 10^{-5}{\rm eV}^2$ and  
$v_u=165{\rm GeV}$.
In following  numerical calculations,
 we take  magnitudes of Yukawa couplings
 to be $0.1\sim 1$. It is found that 
$\alpha_V$ is lower than $10^{-3}$, which is much smaller than
$\alpha_T\simeq 0.032$ in the charged lepton sector.

In the case of the inverted mass hierarchy, 
the situation is different from the case of the normal one.
As seen in $\Delta m^2_{\rm atm}$ of Eq.(\ref{dm2}), 
the sign of $y_0^N$ is opposite against $y_1^N$.
Therefore, $(y_0^N\alpha_S+2y_1^N\alpha_V)$ should be suppressed
compared with $(y_1^N\alpha_V)$ in order to be consistent with
observed ratio $\Delta m^2_{\rm atm}/\Delta m^2_{\rm sol}$.
In terms of the ratio $r$
\begin{equation}
r=\frac{y_1^N\alpha_V}{y_0^N\alpha_S+2y_1^N\alpha_V} \ ,
\end{equation}
we have 
\begin{equation}
\frac{\Delta m_\text{atm}^2}{\Delta m_\text{sol}^2} 
= -r \frac{(y_1^N\alpha_V)^2}{(y_0^N \alpha_S-y_1^N\alpha_V)^2} \ .
\label{r}
\end{equation}
Therefore, we expect $r\sim -100$ for 
$y_0^N \alpha_S \sim -2y_1^N \alpha_V$. 
Then,  we  obtain a typical value:
\begin{eqnarray}
\alpha_V\sim 1.1 \times 10^{-4},
\end{eqnarray}
which is  smaller than the one in the normal hierarchical case 
in Eq.(\ref{normalalpha}).
In the following numerical analyses, we take $r=-100 \sim -10$.

In both cases of normal and inverted mass hierarchies, 
 $\alpha_V$ and  $\alpha_S$ are much smaller than $\alpha_T$. 
Since $\epsilon _i\sim \delta _i\sim \mathcal{O}(\alpha _V^3/\alpha _T^2)$ in Eq. (\ref{epdel}), 
magnitudes of  $\epsilon_i$ and $\delta_i$ are expected to be $10^{-8}$.
 Therefore,  $\epsilon_i$ and $\delta_i$ are negligibly small
 compared with $\alpha_T$,  $\alpha_V$ and $\alpha_S$.

\section{Deviation from the tri-bimaximal mixing}

Let us discuss the deviation from the tri-bimaximal mixing.
In terms of  the  charged lepton mixing matrix and the neutrino one, 
the MNS mixing matrix \cite{MNS} is written as 
\begin{eqnarray}
V_{\rm MNS}
=V_E^\dagger V_{\rm tri} V_{\nu},
\end{eqnarray}
where we have estimated as
\begin{eqnarray}
V_E= \begin{pmatrix}1   & \theta_{12}^e & \epsilon_2 \\ 
                   -\theta_{12}^e    & 1   &\epsilon_1    \\
                   -\epsilon_2  & -\epsilon_1  & 1    \\
 \end{pmatrix},
 \ \
V_{\rm tri}
= \begin{pmatrix}2/\sqrt6   & 1/\sqrt3 & 0 \\ 
                   -1/\sqrt6    & 1/\sqrt3   &-1/\sqrt2    \\
                   -1/\sqrt6  & 1/\sqrt3  & 1/\sqrt2    \\
 \end{pmatrix},\ \ 
V_\nu= \begin{pmatrix}1   & \theta_{12}^\nu & \theta_{13}^\nu \\ 
                   -\theta_{12}^\nu    & 1   &\theta_{23}^\nu    \\
                   -\theta_{13}^\nu  & -\theta_{23}^\nu  & 1    \\
 \end{pmatrix}.
\end{eqnarray}
Then, the deviation from the  tri-bimaximal mixing becomes
\begin{eqnarray}
\delta V_{\rm MNS}
= \begin{pmatrix}\frac{\theta_{12}^e}{\sqrt6}+\frac{\epsilon_2}{\sqrt6}-\frac{\theta_{12}^\nu}{\sqrt3}   & -\frac{\theta_{12}^e}{\sqrt3}-\frac{\epsilon_2}{\sqrt3}+\frac{2\theta_{12}^\nu}{\sqrt6} & \frac{\theta_{12}^e}{\sqrt2}-\frac{\epsilon_2}{\sqrt2}+\frac{2\theta_{13}^\nu}{\sqrt6}+\frac{\theta_{23}^\nu}{\sqrt3} \\ 
                   \frac{2\theta_{12}^e}{\sqrt6}+\frac{\epsilon_1}{\sqrt6}-\frac{\theta_{12}^\nu}{\sqrt3}+\frac{\theta_{13}^\nu}{\sqrt2}    & \frac{\theta_{12}^e}{\sqrt3}-\frac{\epsilon_1}{\sqrt3}+\frac{\theta_{23}^\nu}{\sqrt2}-\frac{\theta_{12}^\nu}{\sqrt6}   &-\frac{\epsilon_1}{\sqrt2}-\frac{\theta_{13}^\nu}{\sqrt6}+\frac{\theta_{23}^\nu}{\sqrt3}   \\
                    \frac{2\epsilon_2}{\sqrt6}-\frac{\epsilon_1}{\sqrt6}-\frac{\theta_{12}^\nu}{\sqrt3}-\frac{\theta_{13}^\nu}{\sqrt2}  & \frac{\epsilon_1}{\sqrt3}+\frac{\epsilon_2}{\sqrt3}-\frac{\theta_{23}^\nu}{\sqrt2}-\frac{\theta_{12}^\nu}{\sqrt6}  & -\frac{\epsilon_1}{\sqrt2}-\frac{\theta_{13}^\nu}{\sqrt6}+\frac{\theta_{23}^\nu}{\sqrt3}    \\
 \end{pmatrix},
\end{eqnarray}
where $V_{\rm MNS}=V_{\rm tri}+\delta V_{\rm MNS}$.
Since magnitudes of  $\epsilon_i$ is found to be  $10^{-8}$,
the charged lepton mass matrix is almost diagonal.
Neglecting $\epsilon_i$ and $\delta_i$, and taking
$\alpha_T\gg \alpha_V\sim \alpha_S$,
neutrino mixing angles are simplified as
\begin{eqnarray}
\hspace{-15mm}
&&\theta_{12}^\nu\approx
\frac{4y_1^Dy_1^N-2y_0^Dy_2^N
+2 y_1^Dy_0^N-2y_0^Dy_{34}^N+y_0^Dy_{35}^N}{3\sqrt{2}y_0^N y_0^D} \ \alpha_T,
\nonumber\\
\hspace{-15mm}
&&\theta_{13}^\nu\approx
-\frac{2y_2^Dy_0^N+3y_0^Dy_{32}^N-3y_0^Dy_{33}^N}
{4\sqrt{3}y_0^D y_1^N}\ \alpha_T,
\nonumber\\
\hspace{-15mm}
&&\theta_{23}^\nu\approx
\frac{y_2^D y_0^N}
{\sqrt{6} y_0^D (2{y_1^N}-{y_0^N})} \ \alpha_T ,
\end{eqnarray}
where these mixing angles are proportional to $\alpha_T$.
Since $\delta_i$ and $\epsilon_i$ are ${\cal O}(10^{-8})$,
the effect of the mixing from the  charged lepton mass matrix
is negligible.
Therefore, the deviation from the tri-bimaximal mixing is 
of ${\cal O}(\alpha_T)$.
Let us estimate typical mixing angles by taking  Yukawa couplings to be 1. 
The typical values of $\alpha$'s are given as
\begin{eqnarray}
\alpha_T\sim\frac{m_\tau}{v_d}\simeq 3.2\times 10^{-2},
\quad 
\alpha_V\sim\frac{\sqrt3 v_u^2}{4\sqrt{\Delta m_{\rm sol}^2}\Lambda}
=5.8\times10^{-4},
\quad 
\alpha_S \sim 7.0\times10^{-4}.
\end{eqnarray}
Therefore, taking 
$\Lambda=2.43\times10^{18} {\rm GeV}$ and using experimental 
values of neutrino mass differences, we obtain
\begin{eqnarray}
\sin^2\theta_{12}\sim0.36,
\quad
\sin^2\theta_{13}\sim4.8\times10^{-6},
\quad
\sin^2\theta_{23}\sim0.48,
\end{eqnarray}
which is a typical prediction in our scheme.


We present the numerical results of neutrino mixing and 
$\alpha_V$ and $\alpha_S$ for both cases of normal and inverted hierarchies. 
Note that we neglect $\epsilon_1$, $\epsilon_2$, 
$\delta_1$, $\delta_2$, which are of order ${\cal}{10^{-8}}$.
The magnitude of $\alpha_T$ is given by the tau mass while 
$\alpha_S$ and $\alpha_V$ are related to neutrino mass squared differences. 
Yukawa couplings are randomly chosen from 
$0.1$ to $1$ with both plus and minus signs. 
Input data of masses and mixing angles are taken in the  region of 
 3$\sigma$ of the experimental data \cite{Threeflavors}:
\begin{eqnarray}
&&\Delta m_{\rm atm}^2=(2.07\sim 2.75)\times 10^{-3} 
{\rm eV}^2 \ ,
\quad \Delta m_{\rm sol}^2= (7.05\sim 8.34) \times 10^{-5} {\rm eV}^2  \ , 
\nonumber \\
&& \sin^2 \theta_{\rm atm}=0.36\sim 0.67 \ ,
\quad  \sin^2 \theta_{\rm sol}=0.25 \sim 0.37  \ , \quad
 \sin^2 \theta_{\rm reactor} \leq 0.056\ .
\end{eqnarray}



In our numerical calculations,  one million random parameter
sets  are produced and only the experimental consistent sets
are plotted in our figures. 
Figure 1 shows our numerical  results  
for the normal hierarchy of neutrino masses. 
 In figures 1 (a) and (b), we plot the allowed region of mixing angles
 on planes of $\sin^2\theta_{12}$-$\sin^22\theta_{23}$ and
$\sin^2\theta_{23}$-$\sin^2\theta_{13}$, respectively,
in the case of $\alpha_V=10^{-4}\sim 10^{-3}$.
The value of $\sin^2 2\theta_{23}$ is larger than $0.97$.
It is also found that the upper bound of $\sin^2\theta_{13}$ is $0.01$.
In figure 1(c), we show the allowed region on 
the $\sin^2\theta_{23}$-$\sin^2\theta_{13}$ plane 
in the case of $\alpha_V=10^{-3}\sim 5\times 10^{-3}$.
It is found that allowed points decrease much more
 in this region of $\alpha_V$.
There are no allowed points   in the region 
of $\alpha_V \geq 5\times 10^{-3}$. 
Thus, $\alpha_V$ is expected to be smaller than ${\cal O}(10^{-3})$.
In figure 1(d),  we plot the allowed region
 on the $\alpha_V$-$\alpha_S$ plane.
 It is found that $\alpha_V \simeq \alpha_S$ 
as expected in Eqs.(\ref{k1}) and (\ref{k2}).

\begin{figure}[tb]
\includegraphics[width=7cm,height=5cm]{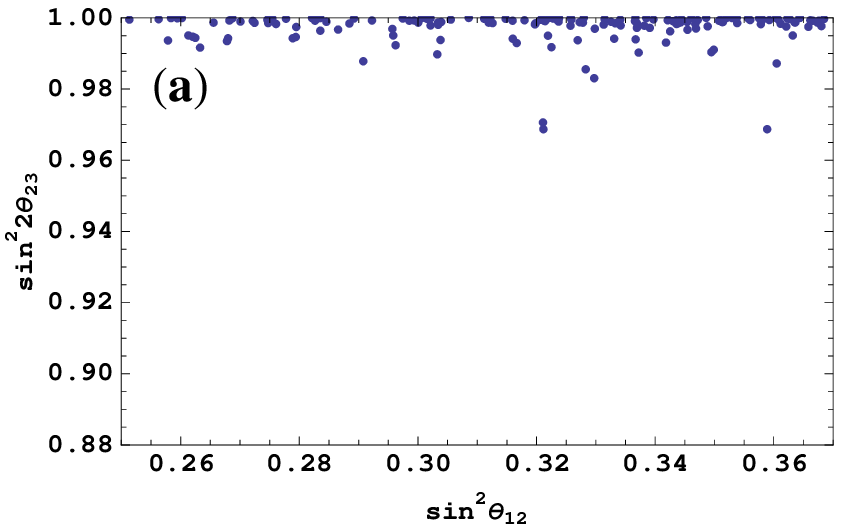}
\hspace{12mm}
\includegraphics[width=7cm,height=5cm]{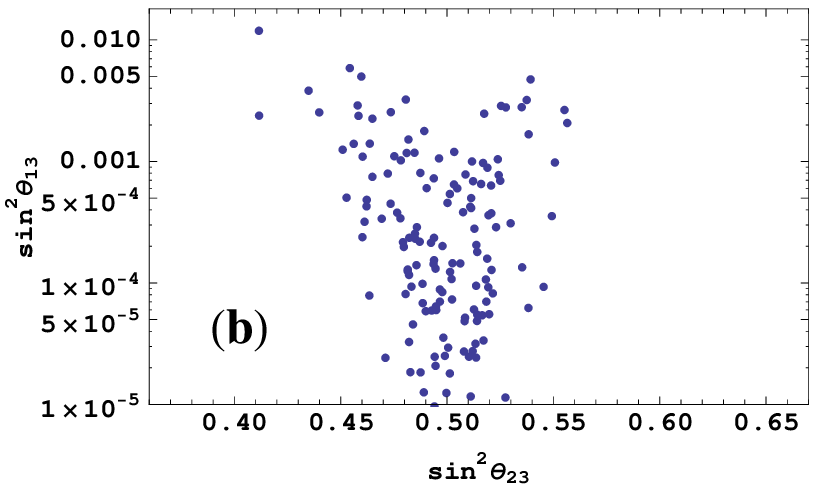}
\end{figure}
\begin{figure}[tb]
\includegraphics[width=7cm,height=5cm]{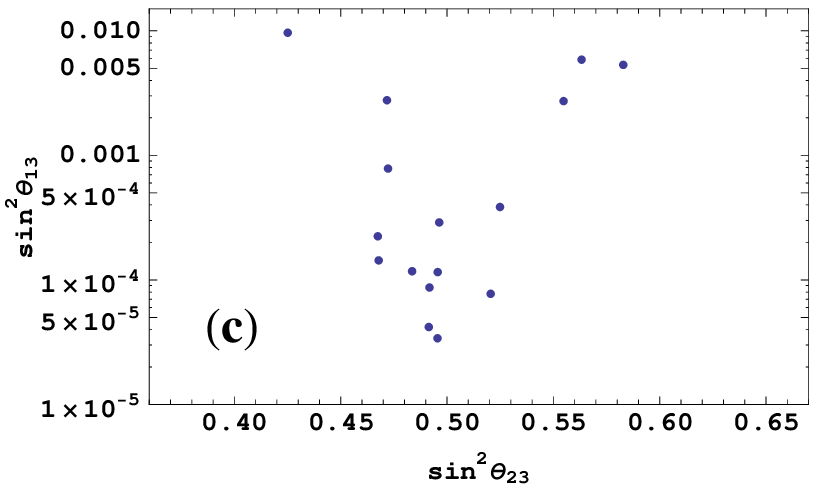}
\hspace{12mm}
\includegraphics[width=7cm,height=5cm]{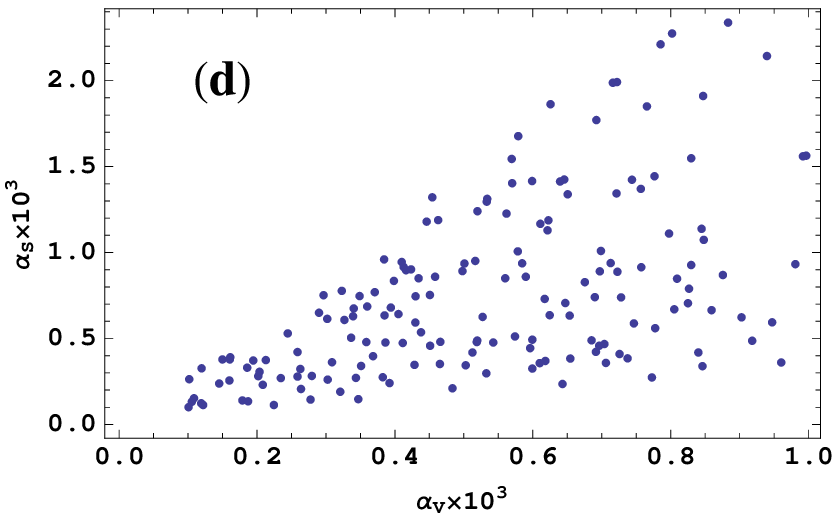}
\caption{Allowed regions on
(a) $\sin^2\theta_{12}$-$\sin^22\theta_{23}$ and 
(b) $\sin^2\theta_{23}$-$\sin^2\theta_{13}$ planes
for $10^{-4}<\alpha_V<10^{-3}$, 
(c) $\sin^2\theta_{23}$-$\sin^2\theta_{13}$ plane
for $10^{-3}<\alpha_V<5\times10^{-3}$, and 
(d) $\alpha_V-\alpha_S$ plane,
 where $1/3<k<3$ is  taken, in the case of the normal hierarchy.}
\end{figure}

Figure  2 shows our numerical results 
for the  inverted hierarchy of neutrino masses. 
The value of $\sin^2 2\theta_{23}$ is larger than $0.96$
as seen in figure 2(a). 
It is also  found that the upper bound of $\sin^2\theta_{13}$ is $0.01$
in figure 2(b). 
These are almost the same result as in the case of the normal hierarchy. 
In figure 2(c), we show the result on the 
$\sin^2\theta_{23}$-$\sin^2\theta_{13}$ plane 
with $\alpha_V=5\times 10^{-4}\sim 10^{-3}$. 
Allowed points decrease considerably 
 in this region of $\alpha_V$. 
There are no allowed points  in the region of
 $\alpha_V \geq 10^{-3}$. 
Thus, $\alpha_V$ should be  smaller than ${\cal O}(5\times 10^{-4})$. 
As in the case of the normal hierarchy, 
$\alpha_V$ and $\alpha_S$ become the same magnitude. 
These values of $\alpha_V$ and  $\alpha_S$ are important parameters 
to estimate the soft SUSY breaking in the next section.

\begin{figure}[tb]
\includegraphics[width=7cm,height=5cm]{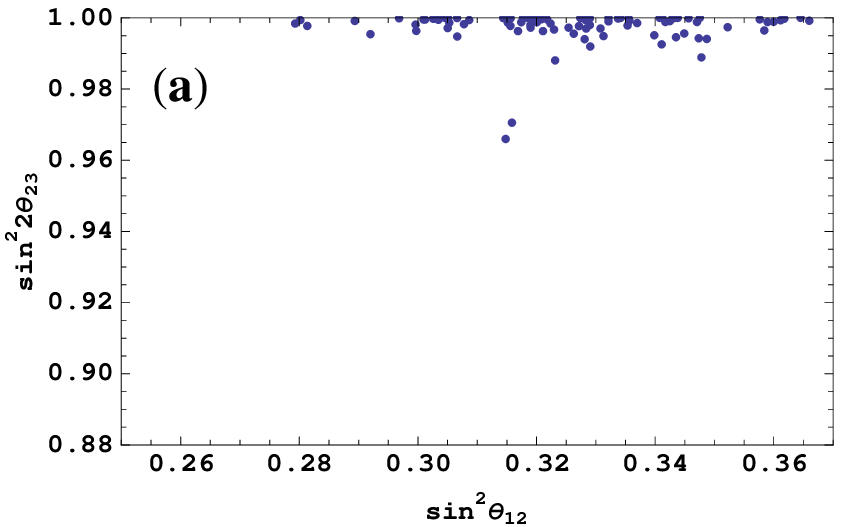}
\hspace{12mm}
\includegraphics[width=7cm,height=5cm]{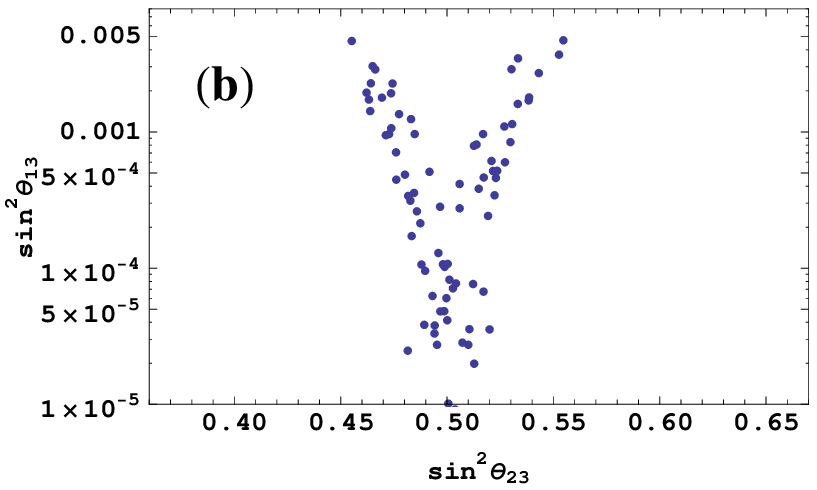}
\end{figure}
\begin{figure}[tb]
\includegraphics[width=7cm,height=5cm]{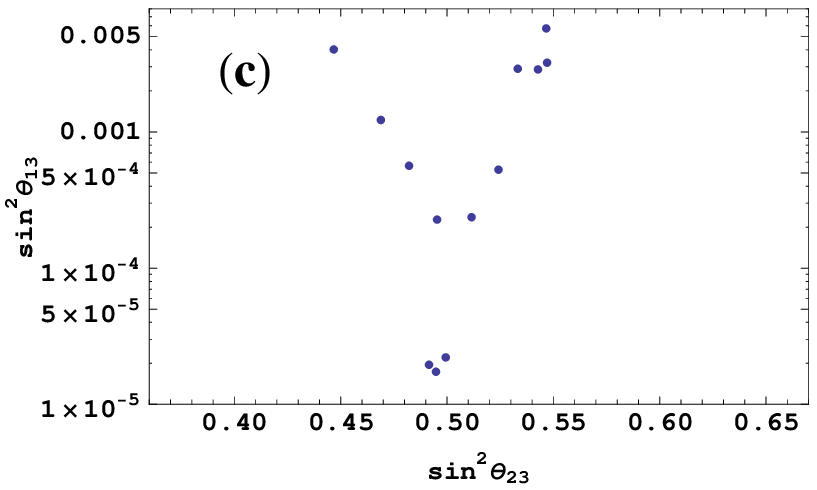}
\hspace{12mm}
\includegraphics[width=7cm,height=5cm]{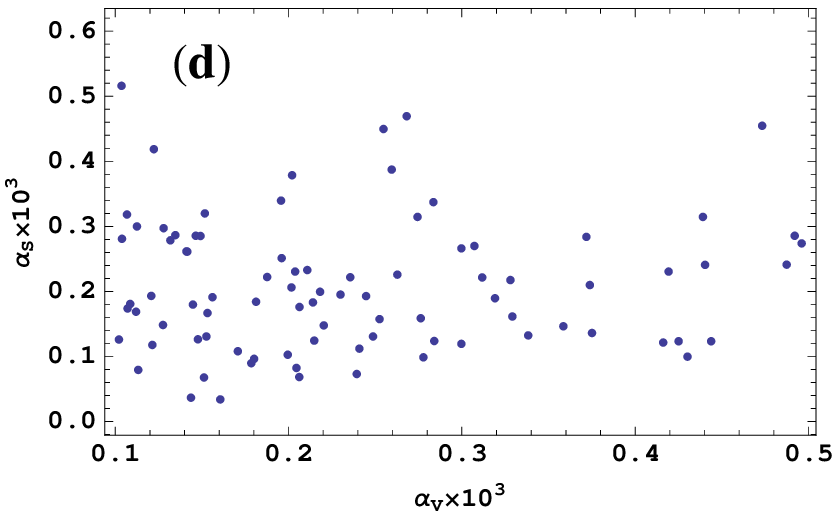}
 \caption{Allowed regions on
(a) $\sin^2\theta_{12}$-$\sin^22\theta_{23}$ and 
(b) $\sin^2\theta_{23}$-$\sin^2\theta_{13}$ planes
for $10^{-4}<\alpha_V<5\times 10^{-4}$, 
(c) $\sin^2\theta_{23}$-$\sin^2\theta_{13}$ plane 
for $5\times 10^{-4}<\alpha_V<10^{-3}$, and 
(d) $\alpha_V-\alpha_S$ plane,
 where $-100<r<10$ is  taken, in the case of the inverted  hierarchy.}
\end{figure}


\section{Soft SUSY breaking terms}

We discuss  soft SUSY breaking terms, i.e. soft slepton masses 
and A-terms, which were discussed in detail the $A_4$ flavor model 
without three right-handed Majorana neutrinos \cite{A4soft}. 
We have obtained the different result in our seesaw type model.


First let us study soft scalar masses.
Within the framework of supergravity theory,
the flavor symmetry $A_4 \times Z_3$ requires the following form 
of K\"ahler potential for left-handed and right-handed leptons 
\begin{eqnarray}
K^{(0)}_{\rm matter}
&=&a(Z,Z^\dagger)(L_e^\dagger L_e+L_\mu^\dagger L_\mu+L_\tau^\dagger L_\tau)
\nonumber\\&&
+b_e(Z,Z^\dagger)R_e^\dagger R_e
+b_\mu(Z,Z^\dagger)R_\mu^\dagger R_\mu
+b_\tau(Z,Z^\dagger)R_\tau^\dagger R_\tau ~ ,
\end{eqnarray}  
at the leading order, 
where $a(Z,Z^\dagger)$ and $b_I(Z,Z^\dagger)$ for
$I = e,\mu,\tau$ are generic functions of moduli fields $Z$.
However, the flavor symmetry $A_4 \times Z_3$ is broken 
to derive the realistic lepton mass matrices 
and such breaking introduces corrections in the K\"ahler potential 
and slepton masses. Because of 
$\langle \phi_{T_2} \rangle, \langle \phi_{T_3} \rangle \ll 
 \langle \phi_{T_1} \rangle$, the 
most important correction terms would be linear terms of 
$\phi_{T_1} $. 
Precisely, the correction terms in the matter K\"ahler potential 
are obtained 
\begin{equation}
\Delta K_{\rm matter} = \frac{\phi_{T_1}}{\Lambda}\left[ a'_1(Z,Z^\dagger)
(2 L_e^\dagger L_e -L_\mu^\dagger L_\mu - L_\tau^\dagger L_\tau)  
 + a'_2(Z,Z^\dagger)(L_\mu^\dagger L_\mu - L_\tau^\dagger L_\tau)
\right ] 
+ h.c.,
\label{eq:K-corr}
\end{equation}
up to ${\cal O}(\tilde \alpha^2)$, 
where $\tilde\alpha$ is the linear combination of 
$\alpha_S$ and $\alpha_V$, 
and $a'_1(Z,Z^\dagger)$ and $a'_2(Z,Z^\dagger)$ are generic functions 
of moduli fields.
All of off-diagonal K\"ahler metric entries for both 
left-handed and right-handed leptons appear at ${\cal O}(\tilde
\alpha^2)$.

Including these corrections, 
the slepton masses are written by 
\begin{eqnarray}
\begin{split}
m_L^2=&
\left(
  \begin{array}{ccc}
m_{L}^2   &  0 &  0 \\ 
0   & m_{L}^2  & 0  \\ 
0   &  0  & m_{L}^2   \\ 
\end{array} \right)
+
m_{3/2}^2
\left(
  \begin{array}{ccc}
 O( \alpha_T)   &  O(\tilde \alpha ^2) & O(\tilde \alpha ^2) \\ 
O(\tilde \alpha ^2)   &  O( \alpha_T)  & O(\tilde \alpha ^2) \\ 
O(\tilde \alpha ^2)   &  O(\tilde \alpha ^2)& O( \alpha_T)   \\ 
\end{array} \right),
\\
m_R^2
=&
\left(
  \begin{array}{ccc}
m_{R_1}^2   &  0 &  0 \\ 
0   &  m_{R_2}^2 & 0  \\ 
0   & 0   & m_{R_3}^2   \\ 
\end{array} \right)
+
m_{3/2}^2
\left(
  \begin{array}{ccc}
 O(\tilde \alpha ^2)   &  O(\lambda^q\tilde \alpha ^2) & O(\lambda^{2q}\tilde \alpha ^2) \\ 
O(\lambda^q\tilde \alpha ^2)   &  O(\tilde \alpha ^2)  & O(\lambda^q\tilde \alpha ^2) \\ 
O(\lambda^{2q}\tilde \alpha ^2)   &  O(\lambda^q\tilde \alpha ^2)& O(\tilde \alpha ^2)   \\ 
\end{array} \right),
\end{split}
\end{eqnarray}
where all of $m_L$ and $m_{R_i}$ for $i=1,2,3$ would be of  
${\cal O}(m_{3/2})$. 
Since  the charged lepton mixing is of 
 ${\cal O}(10^{-8})$, we can neglect its effect.

These forms would be obvious from the flavor symmetry 
$A_4$, that is, three families of left-handed leptons are the 
$A_4$ triplet, while 
right-handed leptons are $A_4$ singlets.
At any rate, it is the prediction of 
the $A_4$ model that 
three families of left-handed slepton masses are almost degenerate.

We have a strong constraint on $(m^2_{L})_{12}$ and 
$(m^2_{R})_{12}$ from FCNC experiments~\cite{FCNCbound}. 
Since $\alpha_S$ and $\alpha_V$ are same order up to $10^{-3}$, 
we can estimate
\begin{equation}
\frac{( m^2_{L})_{12}}{m^2_{\rm SUSY}}\simeq{\cal O}(\tilde\alpha^2)\leq 
{\cal O}(10^{-6}), \qquad \quad
\frac{( m^2_{R})_{12}}{m^2_{\rm SUSY}}\simeq 
 {\cal O}(\lambda^q\tilde \alpha^2) \leq {\cal O}(10^{-7}) ,
\end{equation}
for $m_{\rm SUSY} \sim 100$ GeV, 
where $m_{\rm SUSY}$ denotes the average mass of slepton masses and 
it would be of ${\cal O}(m_{3/2})$.
These predicted values are much smaller than the experimental bound 
${\cal O}(10^{-3})$ \cite{FCNCbound}.

Now, let us examine the mass matrix between left-handed and 
right-handed sleptons, which is generated by the so-called A-terms.
The A-terms are trilinear couplings of two sleptons and one Higgs field 
\cite{A4soft}, i.e.
\begin{equation}
h_{IJ} {R}_I {L}_J H_d =  
h^{(Y)}_{IJ}{R}_I {L}_J H_d  + h^{(K)}_{IJ}{R}_I {L}_J H_d .
\label{eq:A-term}
\end{equation}

The charged lepton mass matrix is diagonalized by
$V_R^\dagger M_l V_L$, where
\begin{eqnarray}
V_R
\sim
\begin{pmatrix}1&\frac{m_e}{m_\mu}\epsilon_2 &\frac{m_e}{m_\tau}\epsilon_1  \\  -\frac{m_e}{m_\mu}\epsilon_2 & 1 &\frac{m_\mu}{m_\tau}\epsilon_2    \\
     -\frac{m_e}{m_\tau}\epsilon_1   & -\frac{m_\mu}{m_\tau}\epsilon_2  & 1    \\
 \end{pmatrix},
 \qquad
V_L
\sim
\begin{pmatrix}1   & \epsilon_1 & \epsilon_2 \\ 
                   -\epsilon_1     & 1   &\epsilon_1     \\
                   -\epsilon_2   & -\epsilon_1   & 1    \\
 \end{pmatrix}.
\end{eqnarray}
In the diagonal basis of the charged lepton mass matrix, 
we estimate the magnitude of 
$\tilde m_{RL}^2\equiv V_R^\dagger  m_{RL}^2V_L$.
By the parallel discussion in \cite{A4soft},
the (2,1) entry of $\tilde m^2_{RL}$ from the second term $h^{K}_{IJ}$ 
 in Eq.(\ref{eq:A-term}) is given as
\begin{equation} 
(\tilde m^2_{RL})_{21} =  {\cal O}(m_\mu\epsilon_1\alpha_T m_{3/2}),
\end{equation}
which gives  $ (\tilde m^2_{RL})_{21}/m^2_{\rm SUSY}={\cal O}(10^{-12})$
 for $m_{\rm SUSY} =100$ GeV.
On the other hand, the first term of (\ref{eq:A-term}) contributes to 
$(\tilde m^2_{RL})_{21}$ as  \cite{A4soft}
\begin{equation}
(\tilde m^2_{RL})_{21} = y_\mu v_d\phi_{T_{2}}m_{3/2}/\Lambda 
\sim m_\mu  \epsilon_1   m_{3/2},
\end{equation}
which gives  $ (\tilde m^2_{RL})_{21}/m^2_{\rm SUSY}={\cal O}(10^{-11})$
 for $m_{\rm SUSY} =100$ GeV.
The predicted value is much smaller than   the FCNC experimental upper bound
 ${\cal O}(10^{-6})$.



\section{Summary}
We have studied the higher order corrections of the flavor symmetry 
breaking in the $A_4$ seesaw model.
 We have discussed possible  higher dimensional mass operators, which
cause the deviation from the tri-bimaximal mixing.
We have found the magnitude of  deviation is dominated by
the VEV of $\phi_{T_1}$, 
which is determined by the tau lepton mass.

The model has $6$ Yukawa couplings 
 ($y_{0}^{e},~y_{0}^{\mu},~y_{0}^{\tau}, ~y_{0}^D, ~y_{0}^N, ~y_{1}^N$)
and $3$ independent VEV's devided by the scale factor 
$\Lambda$, ($\alpha_T,~\alpha_V,~\alpha_S$) at the leading order.
In order to estimate the deviation from the tri-bimaximal mixing,
 we have discussed  higher dimensional mass operators, in which
additional $11$ Yukawa couplings and $3$ VEV parameters appear.
Ratios of charged lepton masses are almost determined by the leading order
 Yukawa couplings as $m_e/m_\tau \propto y_0^e/y_0^\tau$, 
$m_\mu/m_\tau\propto y_0^\mu/y_0^\tau$. Neutrino mass ratios  are also determined
by the leading order Yukawa  couplings
 $y_{0}^D$, $y_0^N$, $y_1^N$ and $\alpha_S/\alpha_V$.

 Since three shift parameters for alignment 
($\epsilon_1=\epsilon_2,~\delta_1,~\delta_2$) are tiny, these effect
is negligibly small both on  mass eigenvalues and flavor mixing angles.
 On the other hand, the deviation from the tri-bimaximal mixing
 depends on additional $7$ Yukawa couplings at the next leading order:
 $y_{1}^D,~y_{2}^D,~y_{2}^N,~y_{32}^N,~y_{33}^N,~y_{34}^N,~y_{35}^N$. 
By varying these Yukawa couplings in the region
$|y_i^{D,N}|=0.1\sim 1$  at random, we can predict 
the deviation from the tri-bimaximal mixing.


We have obtained predictions of lepton mixing angles for both 
normal hierarchy and inverted hierarchy of  neutrino masses.
Since there is no symmetry to suppress the Yukawa couplings, 
we can expect them to be order one. After fixing them, 
mass matrices are determined so that neutrino masses and 
mixing angles can be calculated. 
As our result, the value of $\sin^2 2\theta_{23}$ is larger than $0.96$
and the upper bound of $\sin^2 \theta_{13}$ is $0.01$. 
Therefore, we may expect the Double Chooz experiment  observes
the disappearance of $\overline\nu_e$ 
in the $\overline\nu_e\rightarrow \overline\nu_e$ process.

 It is also found $\alpha_V \sim \alpha_S \leq 10^{-3}$ while 
$\alpha_T \simeq 0.03$. 
In terms of these  values of $\alpha_V$ and  $\alpha_S$, 
we have examined the soft SUSY breaking  in 
slepton masses and A-terms within 
the framework of supergravity theory. Those magnitudes 
are enough suppressed to be consistent with experimental constraints from 
flavor changing neutral current processes. 
This suppression is stronger than that in the case of the effective neutrino 
mass matrix of $A_4$ model, discussed in Ref. \cite{A4soft}. 

\vspace{1cm}
\noindent
{\bf Acknowledgement}

The work of M.T. has been  supported by the
Grant-in-Aid from the  JSPS,  No. 17540243.


\begin{thebibliography}{110}
\bibitem{Threeflavors}
T.~Schwetz, M.~Tortola, and J.W.F.~Valle, New J. Phys.{\bf 10}, 113011 (2008),
arXiv:0808.2016;\\
G.L. Fogli, E. Lisi, A. Marrone, A. Palazzo, and A.M. Rotunno,
Phys. Rev. Lett. {\bf 101} 141801 (2008); arXiv:0806.2649.


\bibitem{fogli}
G.L.~Fogli, E.~Lisi, A.~Marrone,  and A.~Palazzo, 
Prog. Part. Nucl. Phys. {\bf 57}, 742 (2006).

\bibitem{HPS}
P.F. Harrison, D.H. Perkins, and W.G. Scott, 
Phys. Lett. B  {\bf 530}, 167 (2002);\\
P.F. Harrison and W.G. Scott, 
Phys. Lett. B {\bf 535}, 163  (2002).


\bibitem{A4}
E.~Ma and G.~Rajasekaran,
Phys.\ Rev.\ D {\bf 64},  113012 (2001);
\\
E.~Ma,
Mod.\ Phys.\ Lett.\ A {\bf 17}, 2361 (2002);
\\
K.S.~Babu, E.~Ma,  and J.W.F.~Valle,
Phys.\ Lett.\ B {\bf 552}, 207 (2003).

\bibitem{A4-Ma and Rajasekaran}
E.~Ma and G.~Rajasekaran,
Phys.\ Rev.\ D {\bf 64},  113012 (2001),   arXiv:hep-ph/0106291;
\\
E.~Ma,
Mod.\ Phys.\ Lett.\ A {\bf 17}, 2361 (2002).


\bibitem{A4-Babu et al.}
K. S. Babu, E. Ma, and J. W. F. Valle, Phys. Lett.  B{ \bf 552}, 207 (2003), 
 arXiv:hep-ph/0206292.

\bibitem{A4-Hirsch}
M. Hirsch, J.C. Romao, S. Skadhauge, J.W.F. Valle, 
and A. Villanova del Moral, 
Phys. Rev. D {\bf 69}, 093006 (2004),  arXiv:hep-ph/0312265.

\bibitem{A4-Ma}
E. Ma, Phys. Rev. D {\bf 70}, 031901 (2004),  arXiv:hep-ph/0404199.

 \bibitem{Alta1}
 G. Altarelli and F. Feruglio, Nucl. Phys. B {\bf 720}, 64   (2005).

 \bibitem{Alta2}
 G. Altarelli and F. Feruglio, Nucl. Phys. B {\bf 741}, 215 (2006).


\bibitem{A4-Chen et al.}
S.-L. Chen, M. Frigerio, and E. Ma,
  Nucl. Phys. B {\bf 724}, 423 (2005),  arXiv:hep-ph/0504181.

\bibitem{A4-Zee}
A. Zee, Phys. Lett. B {\bf 630}, 58 (2005),  arXiv:hep-ph/0508278.


\bibitem{A4-Adhikary et al.}
B. Adhikary, B. Brahmachari, A. Ghosal, E. Ma, and M. K. Parida, 
Phys. Lett. B {\bf 638}, 345 (2006),  arXiv:hep-ph/0603059.

\bibitem{A4-Valle}
J. W. F. Valle, J. Phys. Conf. Ser. {\bf 53}, 473 (2006),  
arXiv:hep-ph/0608101.

\bibitem{A4-He}
X.-G.~He, Y.-Y.~Keum,  and R.R.~Volkas,
JHEP {\bf 0604}, 039 (2006).

\bibitem{A4-Sawanaka}
E.~Ma, H.~Sawanaka, and M.~Tanimoto, Phys. Lett. B {\bf 641}, 
 301 (2006).

\bibitem{A4-Adhikary and Ghosal}
B. Adhikary and A. Ghosal, Phys. Rev. D {\bf 75}, 073020 (2007),  
                  arXiv:hep-ph/0609193.

\bibitem{A4-Ma(2007)}
 E. Ma, Phys.\ Rev. D {\bf 70},  031901  (2004),
  Phys.\ Rev. D {\bf 72},  037301  (2005),
Mod. Phys. Lett. A {\bf 22}, 101 (2007),  arXiv:hep-ph/0610342.

\bibitem{A4-Altarelli et al.}
G. Altarelli, F. Feruglio, and Y. Lin, Nucl. Phys. B {\bf 775}, 31 (2007), 
 arXiv:hep-ph/0610165.

\bibitem{A4-King}
S. F. King and  M.  Malinsk\'y, Phys. Lett. B {\bf 645}, 351 (2007).

\bibitem{A4-Hirsch et al.}
M. Hirsch, A. S. Joshipura, S. Kaneko, and J.W.F. Valle,
Phys. Rev. Lett. {\bf 99}, 151802 (2007),  arXiv:hep-ph/0703046.

\bibitem{A4-Lavoura}
L. Lavoura and H. K\"uhb\"ock,  
Mod. Phys. Lett. A {\bf 22}, 181 (2007), arXiv:0711.0670.

\bibitem{A4-Honda}
M. Honda and  M. Tanimoto, Prog. Theor. Phys. {\bf 119}, 585 (2008), 
arXiv:0801.0181. 


\bibitem{A4-Bazzocchi}
F. Bazzocchi, S. Kaneko, and S. Morisi, JHEP {\bf 03}, 063 (2008),  
arXiv:0707.3032.

\bibitem{A4-Morisi}
F. Bazzocchi, M. Frigerio, and S.  Morisi,
arXiv:0809.3573.

\bibitem{A4-Lin}
Y. Lin, arXiv:0804.2867.

\bibitem{A4-Hirsch3}
M. Hirsch, S. Morisi, and J.W.F. Valle, arXiv:0810.0121.

\bibitem{A4-Ghosal} 
B. Adhikary and A. Ghosal, Phys. Rev. D {\bf 78}, 073007 (2008), 
arXiv:0803.3582.



\bibitem{Kobayashi:2003fh}
  T.~Kobayashi, J.~Kubo and H.~Terao,
  Phys.\ Lett.\  B {\bf 568}, 83 (2003)
  [arXiv:hep-ph/0303084].


\bibitem{Ko:2007dz}
  P.~Ko, T.~Kobayashi, J.~h.~Park and S.~Raby,
  Phys.\ Rev.\  D {\bf 76}, 035005 (2007)
  [Erratum-ibid.\  D {\bf 76}, 059901 (2007)].


\bibitem{Ishimori:2008ns}
  H.~Ishimori, T.~Kobayashi, H.~Ohki, Y.~Omura, R.~Takahashi and M.~Tanimoto,
  Phys.\ Rev.\  D {\bf 77}, 115005 (2008)
  [arXiv:0803.0796 [hep-ph]].


\bibitem{A4soft}
H. Ishimori, T. Kobayashi, Y. Omura, and M. Tanimoto, JHEP 0812:082 (2008);
arXiv:0807.4625.


\bibitem{FN}
C.D.  Froggatt and H.B. Nielsen,   Nucl. Phys. B {\bf 147}, 277 (1979),

\bibitem{MNS}
Z. Maki, M. Nakagawa,  and S. Sakata,
Prog. Theor. Phys. {\bf 28}, 870  (1962).



\bibitem{FCNCbound}
  F.~Gabbiani, E.~Gabrielli, A.~Masiero and L.~Silvestrini,
  Nucl.\ Phys.\  B {\bf 477}, 321 (1996).


\end{thebibliography}
\end{document}